\documentclass{article}
\usepackage{amssymb,amsmath}
\begin{document}

\title{Polarization from a Strongly Magnetized Accretion Disks:
Asymptotic Wavelength Behaviour}

\author{Gnedin Yu.N.$^{(1)}$, Silant'ev N.A.$^{(1,2)}$, Shternin P.S.$^{(3)}$\\
\\ (1) Central Astronomical Observatory at Pulkovo,
Saint-Petersburg, Russia\\ (2) Instituto National de Astrofisica,
Optica y Electronica, Pue, Mexico\\(3) State Politechnical
University, Saint-Petersburg, Russia}

\maketitle

\begin{abstract}
We calculate the polarization of radiation from thick accretion
disks with vertically averaged global magnetic field. The
polarization arises as a result of the radiation scattering by
free electrons in magnetized plasma of a disk. We consider as a
basic effect the Faraday rotation of polarization plane along the
photon propagation in a magnetized disk. The various models of
optically thick accretion disk with a vertically averaged magnetic
field are considered. The main goal of this paper is to obtain
simple asymptotic formulae for the polarization of radiation when
the Faraday rotation angle $\Psi \gg 1$ at the Thomson optical
length $\tau\approx 1$. The results of our calculation allows us
to estimate the magnetic field magnitude near the marginally
stable orbit region of a black hole via the data of polarimetric
observations including the expected future X-ray polarimetric
observations. The wavelength dependence of polarization is
strongly dependent on various models of an accretion disks and
thus allows to choose a real model from the polarization data.
\end{abstract}

\section{Introduction}
An accretion disk is one of the basic elements for structure of
accreting flows around the compact objects - neutron stars, black
holes, quasars (QSO) and active galactic nuclei (AGN). Due to the
absence of the axial symmetry with respect to a line of sight, the
total integrated radiation emerging from accretion disks should be
polarized via the electron scattering process. The presence of a
magnetic field gives a new effect provided the Faraday rotation of
a polarization plane along a photon mean free path in scattering
medium. Namely, a nontrivial wavelength dependence of polarization
arises when the Faraday rotation angle $\Psi$ at the Thomson
optical depth $\tau$ (see Gnedin and Silant'ev, 1997)

\begin{equation}\label{E_Psi}
  \Psi=0.4 \left(\frac{\lambda}{1\mu m}\right)^2\left(\frac{B}{1
  G}\right)\tau \cos \theta
\end{equation}
is sufficiently large. Here, $\lambda$ is the radiation wavelength
and $\theta$ is the angle between the line of sight and magnetic
field $B$.

The key point is a solution of the Milne problem in the case of
magnetized atmosphere. This problem was considered by many
authors, who obtained the numerical solutions (Silant'ev, 1994,
2002, Agol and Blaes, 1996, Agol et al., 1998, Shternin et al.
2003).

The main goal of this paper is to obtain the analytical asymptotic
formulae for the polarization of radiation and to use them for
analysis of various models of an accretion disk with a global
magnetic field. Here, it is more convenient to use the simple
approximate formulae for the Stokes parameters of emerging
radiation from magnetized, plane-parallel optically thick
atmosphere, obtained by Silant'ev, 2002.

\section{The Degree of Polarization from Optically Thick Accretion
Disks with a Strong Magnetic Field}

Silant'ev, 2002, obtained the quite simple analytical approximate
formulae for the degree of polarization of radiation emerging from
optically thick accretion disk which for considered case
vertically averaged magnetic field acquires the form

\begin{equation}\label{E_SilantForm}
  p_l(\mu)=\frac{1-g}{1+g}\cdot\frac{1-\mu^2}{J(\mu)}\cdot
  \frac{1}{\sqrt{(1-k\mu)^2+\delta^2\mu^2}},
\end{equation}
where $\mu=\cos\theta$, $\theta$ is the angle between the magnetic
field directed along the outward normal (z-axis) and the radiation
emerging from a disk. The function $J(\mu)$ describes the angular
distribution of emerging radiation. This function and the
numerical parameters $g$ and $k$ are tabulated by Silant'ev, 2002.
The parameter of the Faraday depolarization can be introduced from
Eq.(\ref{E_Psi})

\begin{equation}\label{E_delta}
   \delta= 0.8\left(\frac{\lambda}{1\mu
  m}\right)^2\left(\frac{B}{1G}\right).
\end{equation}

We consider the case of dominant electron scattering process
inside an accretion disk ($k=0$, $g=0.83255$).

Photons escape the optically thick disk basically from the surface
layer with $\tau\approx 1$. If the Faraday rotation angle $\Psi$
corresponding to this optical length becomes greater than unity,
then the emerging radiation will be depolarized as a result of the
summarizing of radiation fluxes with very different angles of
Faraday's rotation. Only for directions that are perpendicular to
the vertical magnetic field the Faraday rotation angle is too
small to yield depolarization effect. Certainly, the diffusion of
radiation in the inner parts of a disk depolarizes it even in the
absence of magnetic field because of multiple scattering of
photons. The Faraday rotation only increases the depolarization
process. It means that the polarization of outgoing radiation
acquires the peak-like angular dependence with its maximum for the
direction perpendicular to magnetic field. The sharpness of the
peak increases with increasing magnetic field magnitude. The main
region of allowed angles appears to be $\sim 1/\delta$.

Another the very important feature characterizing the polarized
radiation is the wavelength dependence of polarization degree that
is strongly different from that for Thomson's scattering.

For strong magnetic field magnitude (or large wavelength) when
$\delta\mu\gg 1$ the simple asymptotic formulae takes place

\begin{equation}\label{E_plasymp}
  p_l(B)\approx \frac{p_l(B=0)}{\delta\mu},
\end{equation}
where $p_l(B=0)$ is the classical Chandrasekhar-Sobolev
polarization due to pure electron scattering without magnetic
field.

Silant'ev, 2002, has considered also the other forms of magnetic
field distributions, namely, pure radial and toroidal magnetic
fields. In both cases the Eqs. (\ref{E_SilantForm}) and
(\ref{E_plasymp}) take the same form with changing parameter
$\delta\mu$ to $\delta(1-\mu^2)^{1/2}$. For chaotic magnetic field
the analytical asymptotic dependence does not radically change,
i.e. $p_l(B)\sim 1/\delta$.

Here we restrict ourselves to the case of an accretion disk with
vertically averaged magnetic field. Others distributions of a
magnetic field will be considered in separate works.

\section{Models of Thick Accretion Disks with Vertically Averaged Magnetic Field}

First of all we start with a short review of present models of an
accretion disk with vertically averaged magnetic field (see, for
example, Campbell, 1997, Casse and Keppens, 2002, Li, 2002, Wang
et al., 2003, Romanova et al., 2003, Pariev et al., 2003, Turner
et al., 2003.

Campbell, 1997, in his book "Magnetohydrodynamics in binary stars"
considered the case of a simple dipole magnetic field. For radial
distances from gravitation centre $r\gg z$ he presented the
magnetic field distribution as

\begin{equation}\label{E_CampbField}
  B_z=\frac{1}{2}B_p\left(\frac{R_p}{r}\right)^3, B_r\sim
  \frac{zB_z}{r}.
\end{equation}

The processes of spin-up and spin-down of magnetized stars with
accretion disks and outflows are considered by Romanova et al.,
2003. They has made three-dimensional simulations of disk
accretion to an inclined dipole. It was found the radial
dependence of vertically averaged magnetic field as $B_z\sim
r^{-5/4}$. Using their result one can estimate the vertical
magnetic field magnitude as

\begin{equation}\label{E_BzRom}
  B_z\approx 2.4\cdot 10^3
  \left(\frac{10^8M_\odot}{M}\right)\left(\frac{\dot{M}}{10^{26}g/s}\right)
  \left(\frac{r}{10R_g}\right)^{-5/4}(G),
\end{equation}
where $M$ is the mass of a compact object (a black hole) and
$\dot{M}$ is an accretion rate magnitude, $R_g$ is a gravitational
radius.

Li, 2002, obtained for self-graviting magnetically support disks
the following radial dependence of the vertically averaged
magnetic field

\begin{equation}\label{E_BzLi}
  B_z\approx B_r\approx \left(\frac{2\pi G\Sigma
  M}{R_g^2}\right)^{1/2}\left(\frac{R_g}{r}\right),
\end{equation}
where $\Sigma=\dot{M}/2\pi r v_r$ is the surface density of a
disk. If the equipartition condition would be accepted it provides
the following dependence

\begin{equation}\label{E_BzEquip}
  B_z\approx
  10^2\left(\frac{10^8M_\odot}{M}\right)\left(\frac{\dot{M}}{10^{26}
  g/s}\right)^{1/2}\left(\frac{r}{10R_g}\right)^{-3/2}.
\end{equation}

Casse and Keppens, 2002, developed a model depending on the
central region of a disk, i.e. on $\beta=8\pi
P_{gas}(z=0)/B^2(z=0)$. They obtain the following expression for
vertically averaged magnetic field

\begin{equation}\label{E_BzCas}
  B_z\approx
  B_{10}\frac{1}{\sqrt{\beta}}\left(\frac{10R_g}{r}\right)^{5/2}.
\end{equation}

Pariev et al., 2003, extended the Shakura-Sunayev approach to
strongly magnetized accretion disk model. They assume that the
radial dependence of the vertically averaged magnetic field in the
disk is described by the power law

\begin{equation}\label{E_BzPariev}
  B_z=B_{10}\left(\frac{r}{10R_g}\right)^{-\eta},
\end{equation}
where $B_{10}$ is the strength of the magnetic field at $10R_g$
and $\eta>0$ is some constant. They made numerical calculations
and presented plots of radial structure and emission spectra from
the disk in the region when it is optically thick for four choices
of basic parameters: $\eta=5/4$, $B_{10}=3\cdot10^3 G$; $\eta=1$,
$B_{10}=3\cdot10^3 G$; $\eta=1.4$, $B_{10}=5\cdot10^3 G$;
$\eta=5/4$, $B_{10}=700 G$ and for the mass of supermassive black
hole $M=10^8M_\odot$.

Liu et al., 2003, developed also the model that appears quite
close to Pariev et al. model. Liu et al., 2003, obtained the
following expression for the averaged magnetic field

\begin{equation}\label{E_BzLiu}
  B_z=7.2\cdot10^4\alpha_{01}^{-9/20}\beta_1^{-1/2}\left(\frac{M}{10^8M_\odot}\right)^{-9/20}
  \left(\frac{\dot{M}}{0.1\dot{M}_E}\right)^{2/5}r_{10}^{-51/40}.
\end{equation}
Here, $\alpha$ is the Shakura-Sunayev viscosity coefficient and
$\dot{M}_E$ is the Eddington accretion rate,
$\dot{M}_E=1.4\cdot10^{26}(M/10^8M_\odot)gs^{-1}$. One can see
that their radial dependence of the averaged magnetic field is
practically the same as in Pariev et al. model with $\eta=5/4$.

Turner et al., 2003, developed the model of radiation pressure
supported accretion disk. In their model the vertically averaged
magnetic field was presented by the relation

\begin{equation}\label{E_BzTurn}
  B_z\gtrsim
  10^8\left(\frac{M_\odot}{M}\right)^{1/2}\left(\frac{r}{R_g}\right)^{-\eta},
\end{equation}
where $\eta=3/4$.

The magnetic field was also included in the standard models of an
accretion disk developed by Shakura and Sunayev 1973 and by
Narayan an Yi, 1995 (ADAF model), without detail specification of
its geometry. For these both models the index $\eta=5/4$. For
example, Shakura and Sunayev gave the following expression for the
magnetic field strength

\begin{equation}\label{E_BzShak}
  B_z=B_{10}\left(\frac{10R_g}{r}\right)^{51/40}\lesssim6\cdot10^7
  \left(\frac{M}{M_\odot}\right)^{-17/20}\left(\frac{\dot{M}}{10^{17}g/s}\right)^{2/5}
  \left(\frac{10R_g}{r}\right)^{51/40}.
\end{equation}
It means that $B_{10}\equiv B_{10}(M,\dot{M})$ and
$\eta=51/40\approx5/4$.

The effective temperature of a disk is determined by

\begin{equation}\label{E_StephBoltz}
  \sigma T_e^4(r)=\frac{3GM\dot{M}}{8\pi
r^3}\left(1-\sqrt{\frac{3R_g}{r}}\right).
\end{equation}

The disk radiates as a black body. It means that the peak spectral
wavelength is $\lambda=0.29/T_e$. The last expression allows to
get the relation between radial distance from a compact object and
the peak spectral wavelength corresponding to this distance

\begin{equation}\label{E_RadDist}
  \frac{r}{10R_g}=1.3\cdot10^2\left(\frac{\lambda_m}{1\mu
  m}\right)^2
  M_8^{1/3}\left(\frac{\dot{M}}{\dot{M}_E}\right)^{1/3}.
\end{equation}

As a result one obtains

\begin{equation}\label{E_delta_res}
  \delta=0.8\cdot(1.3\cdot10^2)^{-\eta}\left(\frac{\lambda_m}{1\mu
  m}\right)^\frac{6-4\eta}{3}
  M_8^{\eta/3}\left(\frac{\dot{M}}{\dot{M}_E}\right)^{-\eta/3}B_{10}.
\end{equation}

The Eq.(\ref{E_delta_res}) can also be rewritten in the form

\begin{equation}\label{E_delta_res1}
  \delta=0.8\cdot(0.92)^{\eta}\cdot 10^{-13\eta/3}\left(\frac{\lambda_m}{1\mu
  m}\right)^\frac{6-4\eta}{3}
  \left(\frac{M}{M_\odot}\right)^{2\eta/3}
  \left(\frac{\dot{M}}{10^{17}g/s}\right)^{-\eta/3}B_{10}.
\end{equation}

The strong depolarization takes place only for those wavelengths
which yield $\delta>1$.

\section{The Faraday Depolarization Parameter $\delta$ for Various
Disk Models}

We calculate now the value of the Faraday depolarization parameter
$\delta$ for various models of an accretion disk. We are starting
with the situation of an accretion disk in close X-ray binaries
and consider two cases of magnetic field power law: $\eta=3$
(dipole field) and more popular case when $\eta=5/4$. For $\eta=3$
one obtains from Eq.(\ref{E_delta_res1})

\begin{equation}\label{E_delta3}
  \delta=0.5\cdot 10^{-11}
  \left(\frac{M}{10M_\odot}\right)^2
  \left(\frac{\dot{M}}{10^{17}g/s}\right)^{-1}
  \left(\frac{\lambda_m}{1\mu  m}\right)^{-2}
  B_{10}.
\end{equation}

For optical (V band) photons $\delta\approx 2\cdot10^{-11}
\left(\frac{M}{10M_\odot}\right)^2
\left(\dot{M}/10^{17}g/s\right)^{-1} B_{10}$. It means that for
depolarization of optical photons too much magnetic strength
$B>10^{11} (G)$ is required. For X-rays with the energy $E=1keV$
($\lambda=1.2$\AA) the magnetic field strength $B\gtrsim10^4 (G)$
provides the depolarization of X-ray photons from optically thick
accretion disk.

For $\eta=5/4$ and for V-band of radiation the depolarization
parameter becomes equal to
$\delta=1.35\cdot10^{-5}\left(\frac{M}{10M_\odot}\right)^{5/6}
\left(\dot{M}/10^{17}g/s\right)^{-5/12} B_{10}$, and magnetic
field strength at the inner radius of a disk $B>10^5 (G)$ is quite
sufficient for strong depolarization. The result of calculation
parameter $\delta$ for other values of the index $\eta$ are
presented at Table 1 (V-band photons) and Table 2 (X-ray photons
with the energy 1keV). The most surprising result arises when
$\eta=3/2$. In this case the depolarization parameter $\delta$
does not depend on the wavelength (energy) of photons from a disk.
It means that the polarization will not depend on the wavelength
(energy) of photons as it happens in the classical
Sobolev-Chandrasekhar case. However, the value of polarization
will be less due to the depolarization effect.

One can predict also that, if $\eta>3/2$ the hard energy photons
will be tested by most strong depolarization than soft energy
photons, and the polarization spectrum will drop in the hard
energy region. On the other way, if $\eta<3/2$, the soft energy
photons will be stronger depolarized and the spectrum will drop in
the soft energy region.

\section{The Asymptotic Wavelength Dependence of Polarization\\ from
the Magnetized Standard Accretion Disk}

Now we present the asymptotic formulae for the polarization degree
using the result of analytical calculations by Silant'ev, 2002.
(see Eqs.(\ref{E_SilantForm}) and (\ref{E_plasymp})).

According to Eq.(\ref{E_plasymp}) the asymptotic wavelength
dependence of the optical polarization is determined by

\begin{equation}\label{E_pl1}
  p_l\approx \frac{p_l(B=0)}{\delta\mu}\sim
  \frac{p_l(B=0)}{\mu}10^{2\eta}
  \left(\frac{\lambda}{1\mu m}\right)^\frac{4\eta-6}{3}
  \left(\frac{\dot{M}}{\dot{M}_E}\right)^{\eta/3}/M_8^{\eta/3}
  B_{10}\sim \lambda^\frac{4\eta-6}{3}/\mu,
\end{equation}

or

\begin{equation}\label{E_pl2}
  p_l\sim
  \frac{p_l(B=0)}{\mu}10^{13\eta/3}
  \left(\frac{\dot{M}}{10^{17}g/s}\right)^{\eta/3}
  \left(\frac{\lambda}{1\mu m}\right)^\frac{4\eta-6}{3}
  /\left(\frac{M}{M_\odot}\right)^{2\eta/3}
  B_{10}\sim \lambda^\frac{4\eta-6}{3}/\mu.
\end{equation}

Thus the asymptotic wavelength behaviour of the polarization
allows to derive the value of the index of the power-law radial
dependence of the vertically averaged magnetic field for the
standard magnetized accretion disk. The results of the
calculations for the various models of the standard magnetized
disk are presented in the Table 3.

The basic result from the Table 3 is the radical change of the
wavelength dependence of the polarization at the region of the
index values $\eta=1.5$. For the more strong radial dependence of
the magnetic field ($\eta>3/2$) the polarization is increasing
with the increase of the wavelength. For the more weak radial
dependence ($\eta<3/2$) the polarization is increasing in the
short wavelength region, i.e. $p_l\sim \lambda^{-n}$, where
$n=f(\eta)$. It means that for standard magnetized accretion disk
the only fact of the qualitative wavelength dependence of
polarization (increase or decrease the wavelength) gives an
evidence of the radial dependence of the magnetic field inside an
accretion disk.

\section{Summary and Conclusions}

We presented the formulae for the asymptotic wavelength dependence
of the polarization of radiation from thick accretion disks with
vertically averaged magnetic field. The polarization arises as a
result of the scattering of radiation by electrons in magnetized
plasma of a disk. The effect of the Faraday rotation of
polarization plane in a magnetized disk is taken into account. We
considered the different models of the thick accretion disk with
the vertically averaged magnetic field. The asymptotic wavelength
behaviour of the net polarization displays very strongly the
power-law radial dependence of the vertical magnetic field. It
appears that for more strong dependence (the index of the
power-law radial dependence $\eta>1.5$) the hard energy photon
will be tested by more strong depolarization than soft photons.
For more smooth radial dependence of the magnetic field the soft
energy photons will be stronger depolarized and spectrum of
polarized radiation will drop in the soft energy region. This fact
displays the qualitative estimation the power-law radial
dependence of the magnetic field using only the simple
observations of the super scale dependence of polarization respect
to wavelength of photon (increase or decrease depending on the
wavelength).

\section{Acknowledgements}

One of the authors (P.S.S.) is grateful to the Dynasty Foundation
and ICFPM for financial support.

This work was supported by the RFBR grant 03-02-17223, the Program
of the Presidium of RAN "Nonstationary Phenomena in Astronomy",
the Program of the Department of Physical Sciences of RAN "The
Extended Structure...", and by the Program of Russian Education
and Science Department.

\newpage

\begin{center}
Table 1.\\
 The values of the magnetic field strengths $B_{10}$
 for various magnitudes of the power law index of the magnetic field distributions in the
 standard accretion disk for V-band photons, if $\delta>1$.\\

\begin{tabular}{|c|c|c|c|c|c|c|c|c|}
  \hline
  $\eta$ & 3 & 5/2 & 3/2 & 1.4 & 5/4 & 1 & 3/4 & 0 \\ \hline
  $B_{10}>$ & $10^{11}$ & $1.2\cdot10^9$ & $5\cdot10^6$ & $2.5\cdot10^5$ &
  $10^5$ &$10^4$ &$5\cdot10^3$ &$5$ \\ \hline
\end{tabular}

\bigskip

Table 2.\\
 The values of the magnetic field strengths $B_{10}$
 for various magnitudes of the power law index of the magnetic field distributions in the
 standard accretion disk for X-ray photons ($E_x\approx 1keV$), if $\delta>1$.\\

\begin{tabular}{|c|c|c|c|c|c|c|c|c|}
  \hline
  $\eta$ & 3 & 5/2 & 3/2 & 1.4 & 5/4 & 1 & 3/4 & 0 \\ \hline
  $B_{10}>$ & $10^4$ & $1.3\cdot10^4$ & $5\cdot10^6$ & $8\cdot10^5$ &
  $1.2\cdot10^6$ &$3\cdot10^6$ &$6\cdot10^6$ &$10^8$ \\ \hline
\end{tabular}

\bigskip

Table 3.\\ The asymptotic waveleghth dependence of the
polarization for the standard magnetized accretion disk.

\begin{tabular}{|c|c|c|c|c|c|c|c|c|}
  \hline
  $\eta$ & 3 & 5/2 & 3/2 & 1.4 & 5/4 & 1 & 3/4 & 0 \\ \hline
  $p_l(\lambda)\sim$ & $\lambda^2$ & $\lambda^{4/3}$ & $\lambda^0$ & $\lambda^{-0.133}$ &
  $\lambda^{-1/3}$ & $\lambda^{-2/3}$ & $\lambda^{-1}$ & $\lambda^{-2}$ \\ \hline
\end{tabular}

\end{center}

\end{document}